\documentclass[aps,prb,twocolumn,groupedaddress]{revtex4-2}
\usepackage{graphicx}
\begin{document}

\title{BKT transition observed in magnetic and electric properties of YBa$_2$Cu$_3$O$_{7-\delta}$ single crystals}

\author{V.I. Nizhankovskiy and K. Rogacki}
\email[]{k.rogacki@intibs.pl}
\affiliation{Institute of Low Temperature and Structure Research, Polish Academy of Sciences, ul. Okolna 2, 50-422, Wroclaw, Poland}

\date{\today}

\begin{abstract}
Superconductivity of YBa$_2$Cu$_3$O$_{7-\delta}$ single crystals was investigated in small magnetic fields. In magnetic measurements the superconducting transition for $\textbf{H} \| c$ appears 0.4 K higher than for $\textbf{H} \bot c$. In this temperature range superconductivity is two-dimensional and the total thickness of superconducting layers is about 0.83 of the sample thickness, which is a consequence of the occurrence of the quasi-insulating plane in the unit cell of the crystal structure. Resistivity in the \textit{ab}-plane and along the \textit{c}-axis was measured simultaneously. In these measurements two-dimensional superconductivity was observed in a temperature range of 0.6-0.8 K with the clear signs of the Berezinskii-Kosterlitz-Touless (BKT) transition which occurs approximately 0.15 K below $T_c$, the mean-field transition temperature.
\end{abstract}


\maketitle


\section{Introduction}

The BKT transition (Nobel Prize in 2016) was predicted by Berezinskii \cite{ref1,ref2}, Kosterlitz and Touless \cite{ref3} for two-dimensional (2D) systems (films). If a film becames superfluid at a temperature $T_c$ then the superfluid density, $n_s$, appears at $T_c$ and increases with decreasing temperature if no BKT transition occurs. But if BKT transition takes place at $T_{BKT}$, then at temperatures $T_{BKT} < T < T_c$, the superfluid density remains close to zero due to the creation of pairs of vortices with opposite vorticity. This is possible if the vortex attraction force varies as $1/r$, just as for superfluid He films \cite{Min}. For superconducting films, such behaviour is possible at a distance smaller than the magnetic screening length \begin{equation}
L_s=2\lambda ^2/d ,
\end{equation}
where $\lambda$ is the London penetration depth of bulk material and $d$ is the film thickness \cite{Min,Halperin}. At $T<T_{BKT}$ the vortices are bonded.

One of the most important signatures of BKT transition in superconducting films could be a jump of the exponent $\alpha$ of the I-V characteristics, $V\propto I^{\alpha}$, from $\alpha (T_{BKT})=3$ to $\alpha (T>T_{BKT}) =1$ resulting from a jump of $n_s$ \cite{Halperin}. However, numerous experimental results performed for thin superconducting films of usual and high-temperature (HTSC) superconductors are not convincing because, due to sample inhomogeneity, the BKT transition is blurred \cite{Benfatto,exper} or not observed at all \cite{Rep}. Further theoretical analysis has shown that Eq.(1) which restricts sample dimensions is not sufficient because, due to the boundary conditions at the film edges, the interaction between vorticies turns into a short-range type with near exponential decay \cite{Kogan}. Thus BKT transition will most likely not appear in thin superconducting films of usual superconductor deposited on insulating substrate. Negative edge effects should be suppressed if a film is deposited on a superconducting substrate covered by a thin insulating layer \cite{Kogan}. The ability to eliminate these edge effects in HTSC consisting of a stack of separated superconducting CuO$_2$ planes was not yet examined.

Besides BKT transition that influences the magnetism and electrical conductivity of superconducting films below the mean-field critical Ginzburg-Landau (GL) temperature, $T_c$, fluctuating conductivity of the Aslamazov-Larkin (AL) type exists at $T>T_c$ \cite{AL}. The interpolation formula for conductivity, that combines BKT and AL regimes and describes conductivity in a wide temperature range from $T=T_{BKT}$ to $T \gtrsim T_c$, was proposed in \cite{Halperin}. This formula fits well with our experimental results pointing to the nature of superconductivity near $T_c$.

After the discovery of high-temperature superconductivity, the theory was expanded to explain numerous unusual properties of HTSC in normal and superconducting states. For example, diamagnetism above the critical temperature $T_c$ and in high magnetic fields was observed and explained in \cite{Wang1,Wang2,Oganesyan}. Later the measurements were extended to a wide family of HTSC and high magnetic fields up to 45 T \cite{Wang3}. Until now no studies in small magnetic fields (below 1~Oe) have been reported. In the present work we have investigated the magnetic and resistive properties of YBa$_2$Cu$_3$O$_{7-\delta}$ single crystals in small magnetic fields near T$_c$. Evidence for the BKT transition has been shown in the temperature dependencies of both the critical field and resistivity.

\section{Experiment}

Single crystals of YBa$_2$Cu$_3$O$_{7-\delta}$ (YBCO) were grown in a gold crucible by a conventional self-flux growth method. Then they were annealed in flowing oxygen for 80 hours whilst cooling from 470 $^{\circ}$C to 420 $^{\circ}$C. The single crystals have the shape of a thin plate with the main surface perpendicular to the crystallographic \textit{c}-axis. One large single crystal with a thickness of 21 $\mu$m and a sharp transition to the superconducting state was  broken into 3 pieces: two of them (with a surface $S=3.2\cdot10^{-3}$ cm$^2$) were used for magnetic measurements, one of them ($S=8.8\cdot10^{-3}$ cm$^2$) was used for studying the resistivity.

\subsection{Magnetic measurements}

Most measurements were done with a home-made SQUID magnetometer. This device was constructed for investigation of the quadrupolar magnetic field created by antiferromagnetic Cr$_2$O$_3$ \cite{ref11} and was also used in the measurement of small magnetic moments of antiferromagnetic MnF$_2$ \cite{ref12,ref13}. We avoided the use of superconductors anywhere except for in the pick-up coil and thin-film SQUID sensor. The pick-up coil was placed outside a dewar-insert made from glass. The main cryostat was made from fibreglass and shielded with four thin-wall permalloy tubes. The inner tube had longitudinal and toroidal demagnetizing coils, the next tube had a longitudinal demagnetizing coil. To eliminate trapped flux effects, no superconducting shield was used, even for the SQUID sensor. A residual magnetic field at the sample position, measured with a superconducting lead sphere, did not exceed 1~mOe. The magnetic field was produced by a copper coil. The sample holder was made from five quartz fibres, each 1 mm in diameter. At the end of the holder, aluminium foil and a Cernox-1050 bare chip thermometer were glued. The sample was attached to the foil with Apiezon grease. Good thermal contact between the sample and the thermometer was ensured and thoroughly tested. Once we noticed a parasitic shift in $H_c(T)$ dependence by 0.06 K. Examinations revealed a gap of about 0.3 mm between the thermometer and the aluminium foil. After the thermometer was glued again, the parasitic shift in temperature disappeared.

A zero-field cooling procedure was used, i.e., the sample was cooled down from 95 K to 85 K in zero magnetic field, then the magnetic field was applied and the magnetic moment was measured during slow heating ($dT/dt\simeq 0.12$ K/min) in the fixed field. The magnetization measurements in fields higher than 2~Oe were also performed in a Quantum Design Magnetic Property Measurement System with a Magnet Reset option which allowed us to reduce the trapped field below 0.3~Oe.

\subsection{Resistivity measurements}

\begin{figure}
\centering
 \resizebox{0.9\columnwidth}{!}{%
  \includegraphics{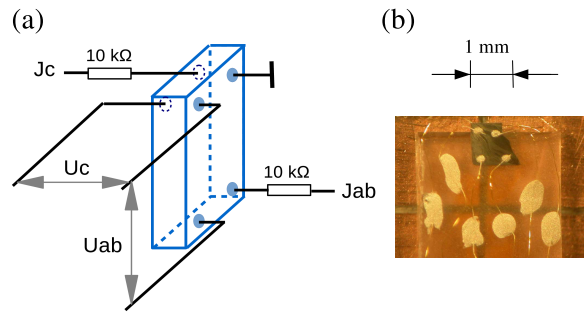}
} \caption{(a) Schematics of the contacts configuration for the resistivity measurements and (b) a photo of the single crystal of YBa$_2$Cu$_3$O$_{7-\delta}$ prepared for investigations.}
\label{fig1}
\end{figure}

A schematic diagram of resistivity measurements is shown in Fig.1(a). Six electrodes were connected to the single crystal for measurements performed in the \textit{ab}-plane and along the \textit{c}-axis. The measurements were done simultaneously with two lock-in amplifiers (Signal Recovery 7225 and 7265) working at different frequencies (83 Hz and 137 Hz). In order to avoid parasitic phase shift, 10~k$\Omega$ resistors (smd type) determining measuring currents were mounted near the sample. The temperature was measured with a Cernox 1050 thermometer. The photo of the YBCO single crystal mounted for the resistivity measurements is shown in Fig.1(b). The crystal was glued to a sapphire plate with GE varnish, and then four contacts (gold wires, 12.7 $\mu$m in diameter) were made on one side of the crystal using a two-component Ag-epoxy (Epoxy Technology, H20E), which was annealed in air at 90 $^{\circ}$C for 80 minutes. Then another two contacts on the other side of the crystal were made in a similar way. In order to obtain low-resistance contacts, the entire assembly was annealed in flowing oxygen at 410~$^{\circ}$C for 12 hours. The contacts made in this way had a resistance of about 1.5~$\Omega$ and were stable over time.

For resistivity measurements, the real temperature of the sample is usually difficult to measure with a high degree of precision. This is especially important for the investigations of I-V characteristics when a large current may overheat the sample. Simultaneous measurements of both $V_{ab}$ and $V_c$ signals allowed us to overcome this difficulty (see below) and removed doubts in the interpretation of the experimental results.

\section{Results}

\subsection{Magnetic moment}

\begin{figure}
\centering
 \resizebox{1\columnwidth}{!}{%
  \includegraphics{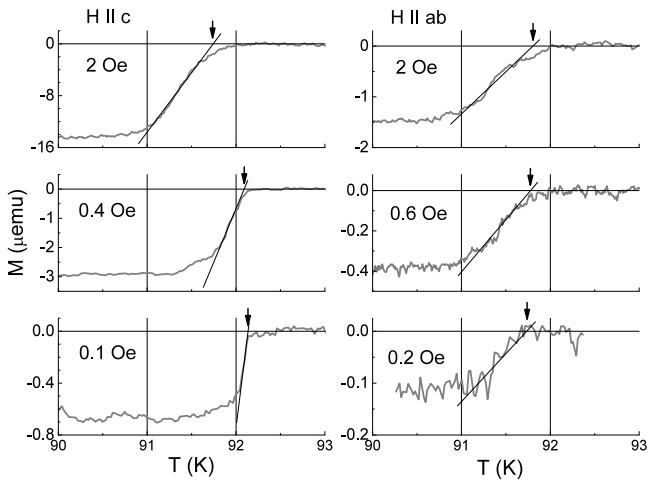}
} \caption{Examples of $M(T)$ records for the single crystal of YBa$_2$Cu$_3$O$_{7-\delta}$. Arrows indicate a corresponding transition temperature, $T_0$.}
\label{fig2}
\end{figure}

Examples of $M(T)$ records for the YBCO single crystal are shown in Fig.2. A significant shift of the superconducting transition to higher temperatures and an increase of its slope was observed for $\textbf{H} \| c$ with decreasing magnetic field from 2~Oe to 0.1~Oe. For $\textbf{H} \| ab$ and in a similar field range, the transition (its temperature, width and slope) remains unchanged. The magnetic moment in the superconducting state (Fig.3) measured for $\textbf{H} \| ab$ was as expected for the single crystal with a volume $v \simeq 7\cdot 10^{-6}$ cm$^{3}$ estimated from its dimensions. The magnetic moment for $\textbf{H} \| c$ was 11.8 times larger than that, which corresponds to the demagnetization factor $n=0.915$.

The superconducting transition temperature, $T_0$, of the single crystal was determined at intersection of the tangent to the $M(T)$ curve with the $M(T)=0$ line (Fig.2 and Fig.4 insert). This definition eliminates the influence of the demagnetization factor of the sample. The width of transition, $\Delta T$, was determined from intersections of the tangent with lines $M(T) = M(85$K) and $M(T) = 0$ (Fig.5 insert). Fig.4 summarizes the $M(T)$ results obtained in fields up to 8~Oe. It is clearly shown that for fields equal to 1.5~Oe and higher, the temperature dependencies of the critical magnetic field of the single crystal for both orientations $\textbf{H} \| c$ and $\textbf{H} \| ab$ converge, but at lower fields they diverge and for $H\rightarrow 0$, $T_0$ for $\textbf{H} \| c$ is 0.4 K higher than for $\textbf{H} \| ab$. Different behaviour of $M(T)$ at the transition, observed for different field orientation, can also be noticed for the field dependence of $\Delta T$ (Fig.5). For $\textbf{H} \| c$, $\Delta T$ increases linearly with a magnetic field below 0.8~Oe, then remains roughly constant, and above 2~Oe increases again with a different slope. For $\textbf{H} \| ab$, the dependence $\Delta T(H)$ is linear for the whole range of magnetic fields.

The two transitions, one for $\textbf{H} \| c$ and the other for $\textbf{H} \| ab$, observed in fields lower than 1.5~Oe at different temperatures, manifest in the shape of the $M(T)$ dependence obtained for $\textbf{H} \| c$, as shown in Fig.6. For clarity, the magnetic moment is normalized to its value at $T=90$~K. It was observed that in fields below 0.8~Oe the transition became two-stage. With increasing temperature, the transition started with a decreasing of the magnetic moment to about 0.83 of its value at $T=90$~K (dashed line in Fig.6), remained at the "0.83 plateau" and then went to zero.

\begin{figure}
\centering
 \resizebox{1\columnwidth}{!}{%
  \includegraphics{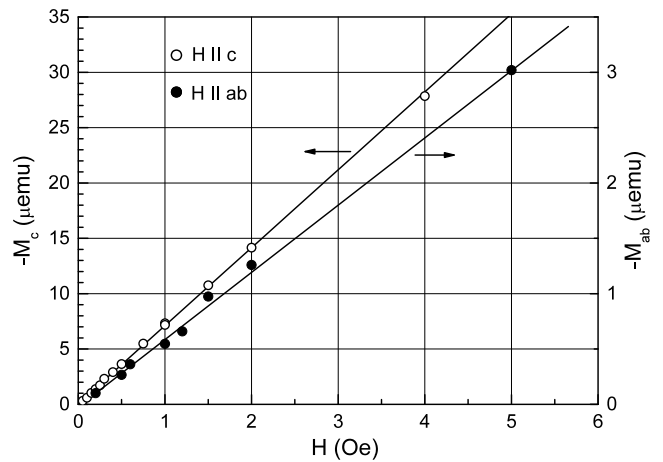}
} \caption{Magnetic moment of the single crystal of YBa$_2$Cu$_3$O$_{7-\delta}$ measured at $T=85$ K for $\textbf{H} \| c$ (open circles) and $\textbf{H} \| ab$ (filled circles).}
\label{fig3}
\end{figure}

\begin{figure}
\centering
 \resizebox{1\columnwidth}{!}{%
  \includegraphics{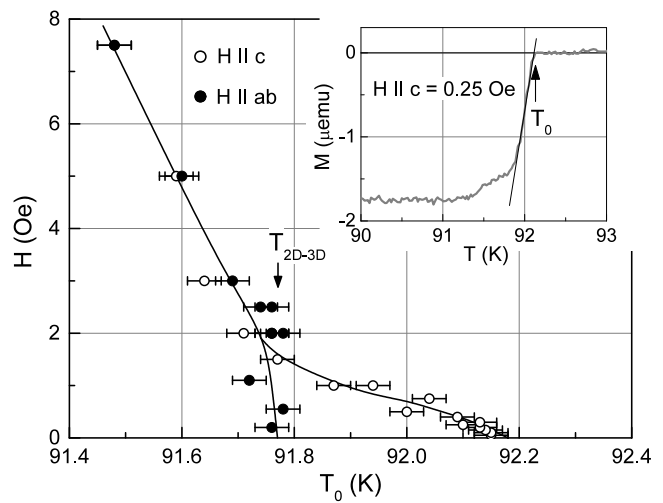}
} \caption{Critical magnetic field of the single crystal of YBa$_2$Cu$_3$O$_{7-\delta}$. Insert shows the definition of the transition temperature $T_0$.}
\label{fig4}
\end{figure}

\begin{figure}
\centering
 \resizebox{1\columnwidth}{!}{%
  \includegraphics{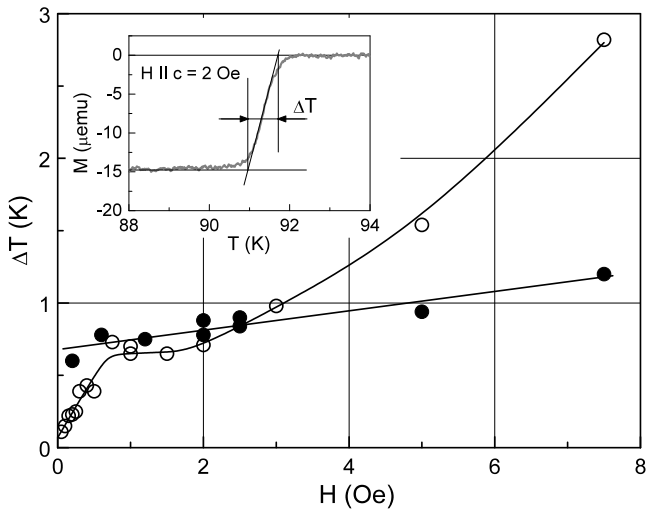}
} \caption{Width of the superconducting transition of the single crystal of YBa$_2$Cu$_3$O$_{7-\delta}$ as a function of the applied magnetic field $\textbf{H} \| c$ (open circles) and $\textbf{H} \| ab$ (filled circles). The insert shows the definition of $\Delta T$.}
\label{fig5}
\end{figure}

\begin{figure}
\centering
 \resizebox{1\columnwidth}{!}{%
  \includegraphics{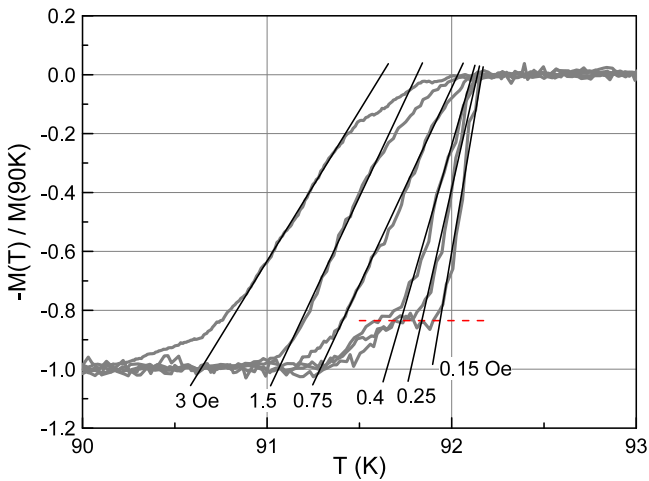}
} \caption{Change in the shape of the temperature dependence of the magnetic moment at the superconducting transition of the single crystal of YBa$_2$Cu$_3$O$_{7-\delta}$ for $\textbf{H} \| c$. The magnetic moment is normalized to its value at $T=90$ K.}
\label{fig6}
\end{figure}

\subsection{Resistivity}
Resistivity in the \textit{ab}-plane (Fig.1) was measured in the Van der Pauw configuration \cite{Pauw} with contacts placed near to, but not at the sample's edges. A correction coefficient of 1.08 for the non-standard geometry of the sample was obtained experimentally using a model sample cut from brass foil.

Resistivity along the \textit{c}-axis was measured using the Schnabel configuration \cite{Sch1,Sch2}. For the ideal case of an infinite plate with point contacts, the resistance is:
\begin{equation}
 R_c=U_c/I_c= \frac{\rho_{ab}}{\pi d}G(\alpha)
 \end{equation}
were $\alpha=(\rho_c/\rho_{ab})^{0.5}\cdot \alpha_0$, $\alpha_0=d/s$, \textit{d} is the plate thickness and \textit{s} is the distance between current and potential contacts. Function $G(\alpha)$ is tabulated in \cite{Sch1}. The derivative $dG/d\alpha \rightarrow 0$ for $\alpha\rightarrow 0$ and $dG/d\alpha \rightarrow 1$ for $\alpha\rightarrow \infty$. Thus for $\rho_c=const$, $R_c$ should diminish with diminishing $\rho_{ab}$ and $R_c(\rho_{ab}=0)=0$. From measurements performed at room temperature, $\rho_{ab}=0.37$ m$\Omega$cm and $R_c=1.35$ m$\Omega$, and taking $d=21$ $\mu$m and $s=0.37$ mm, we got $G=0.024$, $\alpha=0.689$ and $\rho_c/\rho_{ab}\simeq 150$. Taking into account that the contacts were placed near one side of the sample whose maximum size is only three times larger than the distance $s$ between the contacts, the real value of an anisotropy is several times smaller.

If the plate is a rectangle, then the function $G(\alpha)$ may be calculated by the method proposed in \cite{Lug}.
Numerical estimation was done for a rectangle 0.8~mm x 1.2~mm and for contacts placed at its edge. Results for $s=0.37$~mm were $\alpha=0.605$ and $\rho_c/\rho_{ab}\simeq 113$. If \textit{s} is made equal to the minimal distance between contact spots, $s=0.25$~mm, then $\alpha=0.619$ and $\rho_c/\rho_{ab}\simeq 54$. This anisotropy is higher than the maximum value $\rho_c/\rho_{ab}=35$ reported for optimally doped YBCO \cite{Nag}. Larger anisotropy is expected for single crystals of higher quality.

Resistivity in the \textit{ab}-plane and along the \textit{c}-axis of the YBCO single crystal was measured on cooling from room temperature to below $T_c$ and is shown in Fig.7(a). For temperatures above 93 K, $R_c$ is proportional to $\rho_{ab}$, so the anisotropy is constant. A detailed picture of the superconducting transition is shown in Fig.7(b). The data was obtained with slow heating of 0.2 K/min.  Analysing the results when going from the normal to the superconducting state, it is seen that resistivity in the \textit{ab}-plane monotonically diminishes (the width of transition is about 0.30 K), whereas resistance along the \textit{c}-axis starts to increase at the beginning of the transition in the \textit{ab}-plane, goes through a deep minimum when $\rho_{ab}(T)\simeq 0.1 \rho_n$ and becomes zero below $T=T_{2D-3D}$, where the transition along the \textit{c}-axis is completed. The results are qualitatively the same for the two configurations of current and potential leads (for details see Fig.7(b)), and only the depth of the minimum differs. This means that the minimum in $R_c(T)$ curves cannot be ascribed to a non-alignment of contacts or inhomogeneous distribution of current and has a deeper physical meaning, as we will discuss in section IV. Increasing the measuring current from 50 to 400 $\mu$A had no influence on $R_c(T)$ dependence.

Results shown in Fig.7 were obtained during the first week after the sample's contacts were prepared. Measurements performed six weeks later, which allowed the sample to relax and return to a state with homogeneous oxygen distribution, yielded the results presented in Fig.8. The temperature dependence of the resistance along the \textit{c}-axis changed, however some characteristic features remain. As before, $R_c(T)$ begins to grow when $\rho_{ab}(T)$ begins to diminish, shows two maxima and goes to zero at $T=91.9$ K (Fig.8(a)) instead of $T=92.2$ K (Fig.7(b)). For $R_c(T)$, the more pronounced maximum is observed  at $T=92.74$ K where $\rho_{ab}(T)$ reaches zero. When superconductivity in the \textit{ab}-plane is suppressed by a larger measuring current, the increase of $R_c$ shifts to lower temperatures and the more pronounced maximum vanishes (Fig.8(b)). The temperature dependence of the resistivity in the \textit{ab}-plane was measured at several currents from 20 $\mu$A to 5 mA, which allowed us to determine the exponent $\alpha$ of I-V characteristics, $V\propto I^{\alpha (T)}$ (Fig.9(a)), and the critical current at a level $V_{ab}=10$ nV (Fig.9(b)). For these measurements, a current limiting resistor was reduced from 10 k$\Omega$ to 1 k$\Omega$ to allow the use of a sufficiently high current.

\begin{figure}
\centering
 \resizebox{0.9\columnwidth}{!}{%
  \includegraphics{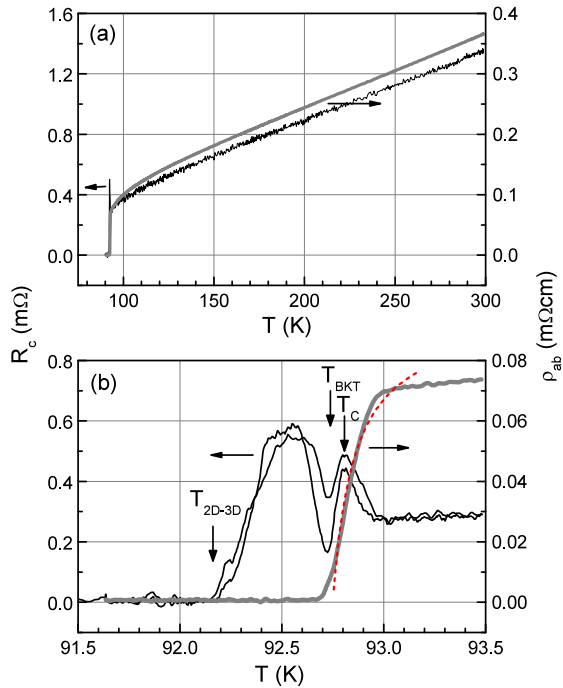}
} \caption{(a) Temperature dependence of the resistivity of the single crystal of YBa$_2$Cu$_3$O$_{7-\delta}$: grey line shows the resistivity in the \textit{ab}-plane, $\rho_{ab}$, at $I_{ab}=50$ $\mu$A; black line shows the resistance along the \textit{c}-axis, $R_c$, at $I_c=200$ $\mu$A. (b) Detailed picture of $\rho_{ab}(T)$ and $R_c(T)$ at the superconducting transition, for $I_{ab}=20$ $\mu$A and $I_c=100$ $\mu$A. Black lines show $R_c(T)$ measured for interchanged current and potential leads. Dashed line in (b) shows fitting of $\rho_{ab}(T)$ to Eq.(4).}
\label{fig7}
\end{figure}

\begin{figure}
\centering
 \resizebox{0.9\columnwidth}{!}{%
  \includegraphics{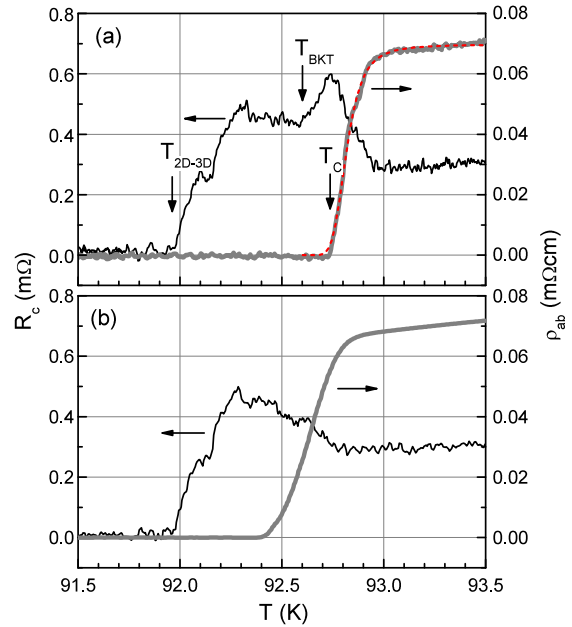}
} \caption{Resistivity at the superconducting transition of the single crystal of YBa$_2$Cu$_3$O$_{7-\delta}$ measured in the \textit{ab}-plane (grey line) and along the \textit{c}-axis (black line) at: (a) $I_c=200$~$\mu$A, $I_{ab}=50$~$\mu$A and (b) $I_c=200$~$\mu$A, $I_{ab}=5$~mA. Dashed line in (a) shows fitting of $\rho_{ab}(T)$ to Eq.(5). These results were obtained 6 weeks later than the results presented in Fig.7.}
\label{fig8}
\end{figure}

\begin{figure}
\centering
 \resizebox{0.8\columnwidth}{!}{%
  \includegraphics{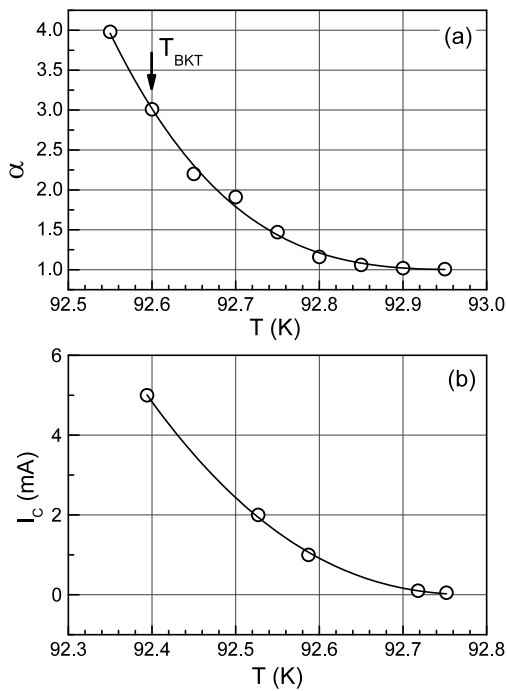}
} \caption{(a) Temperature dependence of the exponent $\alpha$ in I-V characteristics, $V_{ab}\propto I^{\alpha (T)}$, of the single crystal of YBa$_2$Cu$_3$O$_{7-\delta}$. (b) Temperature dependence of the critical current determined at a level $V_{ab}=10$ nV. Solid line means $I_c(mA)=43(92.74-T)^{2.33}=43[{T_c}(1-T/T_c)]^{2.33}$, where $T_c = 92.74$ K (see Fig.8(a)).}
\label{fig9}
\end{figure}

\section{Discussion}

\begin{figure}
\centering
 \resizebox{0.6\columnwidth}{!}{%
  \includegraphics{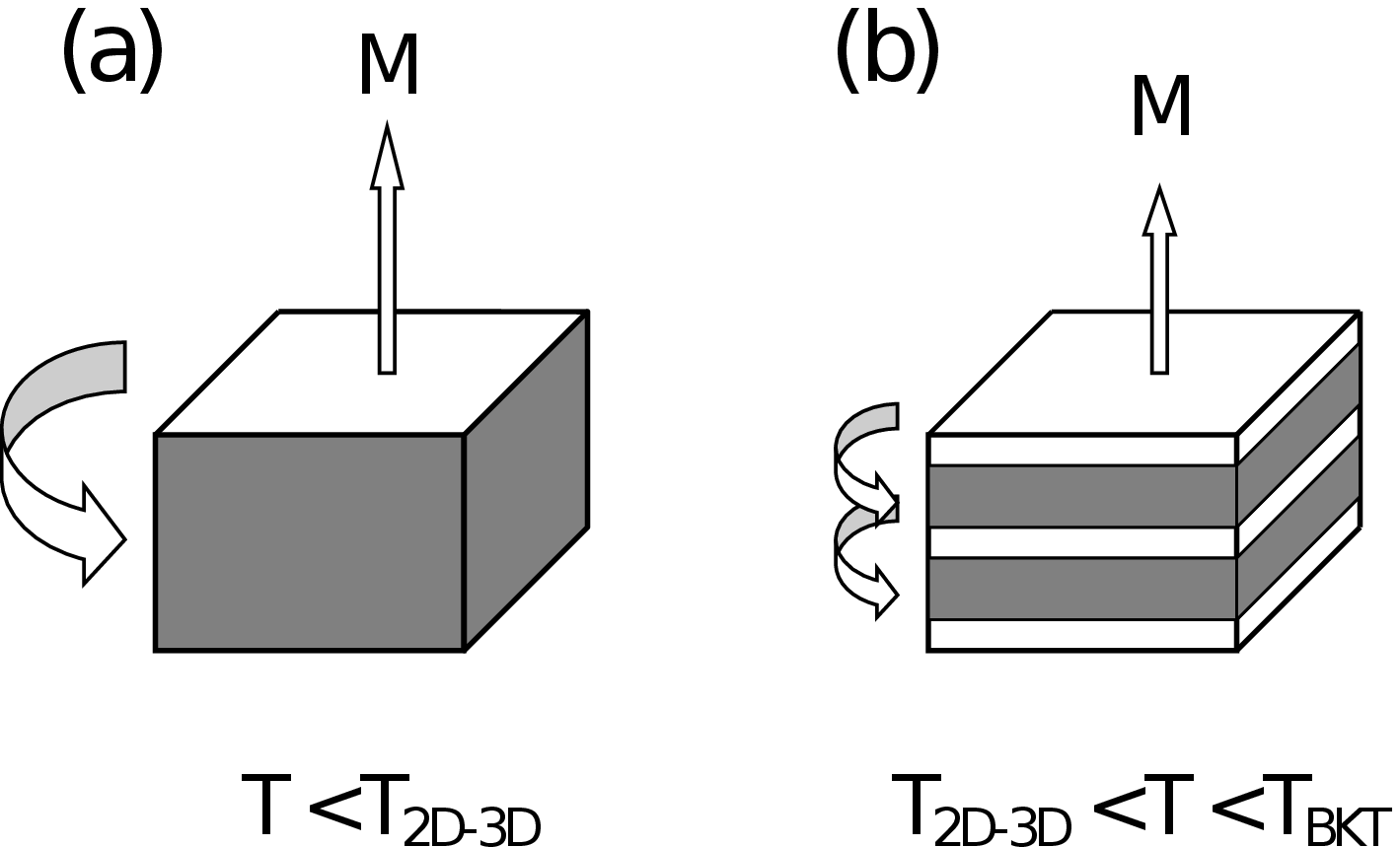}
} \caption{Illustration of the screening current flow in the Meissner state at temperatures from two ranges: (a) $T<T_{2D-3D}$ and (b) $T_{2D-3D}<T<T_{BKT}$. The \textit{c}-axis of the single crystal of YBa$_2$Cu$_3$O$_{7-\delta}$ is oriented parallel to the magnetic field and thus to the magnetic moment, $M$. }
\label{fig10}
\end{figure}

Experimental data for magnetization of the YBCO single crystal (Fig.4) clearly show that it has two different superconducting states. Below $T_{2D-3D}=91.76$~K, YBCO behaves as a usual three-dimensional (3D) superconductor. Above this threshold temperature, its behaviour is different - large diamagnetism remains for $H \| c$, however, it disappears for $H \| ab$. If we identify $T_{2D-3D}$ with the critical temperature $T_c$, then the diamagnetic response above $T_c$ may result from the Gaussian thermal fluctuations of the order parameter $\psi$  \cite{ref14,ref15}. However, in that case the theory does not predict any critical magnetic field above which this diamagnetism should disappear, which is contradictory to our observations, as shown in Fig.4.

Another explanation could be proposed if one takes into account the crystalline structure of YBCO, which is a layered compound with conducting (metallic) layers (blocks of double CuO$_2$ planes with one yttrium layer) separated by buffer layers (blocks of CuO chain with two barium layers) serving as a charge reservoir. Thus, this compound is considered as a quasi 2D system due to the smallness of the coherence length along the $c$-axis, $\xi_c$, with respect to the thickness of the buffer layer. In this case, with temperature decreasing from $T=92.18$~K, the superconductivity appears in CuO$_2$ planes forming a stack of weakly coupled quasi 2D superconducting plates. Further lowering of the temperature will result in the transition to bulk superconductivity at $T_{2D-3D}$. When a magnetic field smaller than the lower critical field, $H_{c1}$, is applied along the \textit{c}-axis at $T=85$~K, the sample acquires a negative magnetic moment due to the surface screening current. For $T<T_{2D-3D}$, this current fills the entire side surface of the sample, as shown schematically in Fig.10(a). When temperature rises above $T_{2D-3D}$, the screening current should split up into many streams along the sides of the 2D superconducting plates (Fig.10(b)). Consequently, if the applied magnetic field is small enough that the screening current does not exceed the critical value, then the $M(T)$ dependence will reveal a double-stage transition to the normal state, as shown in Fig.6. For $H \parallel c<1$~Oe, a small plateau is observed at $M(T)/M(90$K)=0.83, which corresponds to the magnetic moment of the separated quasi 2D superconducting plates. The results presented in Fig.6 allowed us to estimate the thickness of the superconducting plates in one unit cell to be equal to $d_s=0.83\cdot c=0.97$ nm ($c=1.17$ nm is the lattice dimension along the $c$-axis) and, consequently, the thickness of non superconducting layers to be equal to 0.20 nm.

In the Meissner state the screening current flows in a sheet with a thickness of about the penetration depth, $\lambda$. For $H \| c$ (Fig.10), an increase of temperature above $T_{2D-3D}$ does not change the screening current trajectory; it continues to flow along the perimeter surface of the sample, except of the non superconducting buffer layers, as shown in Fig.10(b). For $H \| ab$, the situation is different; instead of flowing along the perimeter surface of the sample (as for $T<T_{2D-3D}$), the screening current will flow along the perimeter surface of each superconducting plate, that is the block of double CuO$_2$ planes with an yttrium layer. For this orientation of the magnetic field ($H \| ab$), the current trajectory will run along the $ab$-plane and in the direction of the $c$-axis. At $T<T_{2D-3D}$ the thickness of the current sheet is about $\lambda_c (T)=\lambda_c (0)(1-T/T_c)^{-1/3}$. Using $\lambda_c (0)\approx 8.7\cdot 10^{-5}$~cm \cite{Lambda_c} and $T_c=92.18$~K we get $\lambda_c(T_{2D-3D})\approx 3.9\cdot 10^{-4}$~cm, for $T_{2D-3D}=91.76$~K, as derived from Fig.4. This value is more than 5 times smaller than the thickness of the sample, that explains the full Meissner state revealed at temperatures just below $T_{2D-3D}$, but is much larger than the thickness of a single superconducting plate ($d_s\simeq1$~nm), which explains the abrupt suppression of the Meissner state at $T_{2D-3D}$ for $H \| ab$ (Fig.4).

Earlier, the thickness of the superconducting plates in several high-temperature superconductors was estimated from the crossing-point phenomenon in magnetization measurements \cite{Schneider}. In numerous experiments it has been observed that the magnetization curves $M(H,T)$ of most HTSC cross at a single point $M^*(T^*)$, which is independent of the magnetic field applied perpendicularly to the superconducting planes. According to the theory \cite{Schneider}, $T^*=T_{BKT}$ and
\begin{equation}
-M^*=\frac{0.52 k_B T_{BKT}}{2d_s\Phi_0},
\end{equation}
where $k_B$ is the Boltzmann constant and $\Phi_0$ is the magnetic flux quantum. It should be noted that in other theoretical works, $d_s$ was assigned to the interlayer distance and the numerical coefficient in Eq.(3) was different \cite{Bul,Kosh,Jun}. In Ref. \cite{Schneider}, the values of $d_s$ were from 0.76 nm for La$_{1.92}$Sr$_{0.08}$CuO$_4$ to 4.4 nm for TlBa$_2$Ca$_2$Cu$_3$O$_{9+\delta}$. The last value is almost 3 times larger than the lattice constant along the $c$-axis which seems to be impossible. The crossing point was not observed for YBCO due to its relatively low anisotropy.

Behaviour of the resistivity of our sample in the normal state (Fig.7 (a)) is typical for optimally doped YBCO \cite{Nag,Shams,Uch}. For $T>93$~K, $R_c(T)$ changes in the same way as $\rho_{ab}(T)$. However, the behaviour of $R_c(T)$ alters at lower temperatures with the beginning of the superconducting transition in the \textit{ab}-plane, and this points to the emergence of different mechanism of conductivity along the \textit{c}-axis. Like in our work, in \cite{Vovk} it was observed that the superconducting transition in the \textit{ab}-plane occurs at higher temperatures than that along the \textit{c}-axis. For our sample, $T_c^{ab} - T_c^c = 0.74$~K, in \cite{Vovk} it is equal to 0.35~K.

Electrical properties of YBCO may be modelled by a stack of conducting plates separated by thin buffer layers. At high temperatures the conductivity along the \textit{c}-axis is provided by single electron transport through the buffer layers between conducting plates. In this case, the behaviour of resistance along the \textit{c}-axis (Fig.7(b)) may be explained in the following way; as the temperature decreases and fluctuating superconductivity in the conducting blocks of double CuO$_2$ planes appears ($\rho_{ab}$ decreases), the number of non-paired electrons in these planes, $n_n$, diminishes which results in an increase of resistance along the \textit{c}-axis. If BKT transition occurs at $T_{BKT}<T_c$, then for temperatures in the range  $T_{BKT}<T<T_c$, the superfluid density $n_s$ remains zero or close to zero, $n_n$ stays roughly constant and $R_c(T)$ does not change or decreases due to more uniform current distribution. Thus a maximum in the $R_c(T)$ dependence appears just above $T_{BKT}$. Below this temperature, $n_s$ rapidly increases, $n_n$ decreases and $R_c$ raises again. At lower temperatures $R_c$ begins to diminish due to Josephson coupling (Cooper pairs tunnelling) between the blocks of double CuO$_2$ planes down to $R_c=0$, which corresponds to the 2D-3D transition. This qualitatively explains the results presented in Fig.7(b). In this figure, arrows correspond to $T_{BKT}=92.72$~K and $T_c=92.81$~K, so the temperature interval in which the BKT vortex state exists is very narrow (0.09~K).

Resistivity measurements were repeated after six weeks and the results shown in Fig.8 confirmed the observed increase of $R_c$, which occurred simultaneously with the decrease of $\rho_{ab}$ when the temperature was lowered. In Fig.8(a), the resistance along the \textit{c}-axis reaches a maximum at $T_c=92.74$~K when $\rho_{ab}=0$. The BKT transition may be assigned to the temperature $T_{BKT} = 92.60$~K, where $R_c$ stops decreasing, so for these results, the temperature interval where the BKT vortex state exists is 0.14 K. Fig.8(b) shows that the superconducting transition in the \textit{ab}-plane is suppressed by a large measuring current and the maximum in $R_c(T)$ is smeared off, as expected.

Exponent $\alpha$ of the I-V characteristics, $V_{ab}\propto I^{\alpha (T)}$, gradually increases from $\alpha =1$ to $\alpha =4$ when the temperature decreases from 92.95 K to 92.55 K (Fig.9(a)). From condition  $\alpha(T_{BKT})=3$ it follows that $T_{BKT}=92.60$ K, which corresponds perfectly with the resistivity results (Fig.8(a)). In the temperature range $T_{2D-3D} <T<T_{BKT}$, $R_c$ is determined by the competition between single-electron transport, which diminishes with lowering temperature, and Cooper pairs tunnelling, which rises \cite{Gray}. Thus, different behaviour of the resistivity can be observed and a quantitative model is necessary to describe the specific $R_c(T)$ dependence. To our knowledge, such a quantitative model has not been presented to date in literature.

An attempt to describe the resistive transition in the ab-plane was performed by using the Aslamazov-Larkin (AL) paraconductivity model with the resulting formula \cite{AL}
\begin{equation}
1/\rho=1/\rho_n	+ \sigma\sqrt{T/T_c -1},
\end{equation}
where $\sigma$ is a constant. This attempt was not successful, as shown in Fig.7(b). Similarly unsuccessful was an attempt to describe the results presented in Fig.8. Thus, the nature of the state above the transition to the superconducting state seems to be different to the fluctuating superconductivity of the AL type, which is based on the appearance of the Cooper pairs above $T_c$ along with local condensation. On the other hand, the temperature dependence of the resistivity in the \textit{ab}-plane, measured at small current (grey line in Fig.8(a)), was well mapped by the formula obtained in the frame of the model which interpolates the BKT behaviour with the GL-type fluctuations \cite{Benfatto}:
\begin{eqnarray}
\frac{\rho}{\rho_n}=\frac{1}{1+(\xi / \xi _0)^2}, \quad T \geq T_{BKT}, \\
\frac{\xi}{\xi _0}=\frac{2}{A}sinh\frac{b}{\sqrt{t}}, \quad t=\frac{T-T_{BKT}}{T_{BKT}} \nonumber,
\end{eqnarray}
where $\xi_0$ is the zero-temperature coherence length and $A$ and $b$ are fitting parameters, which we will discuss below. As shown in Fig. 8(a) (dashed line), Eq.(5) describes the experimental $\rho_{ab}(T)$ very well in the whole region of the transition to the superconducting state. The fitting parameter $b$ is:
\begin{eqnarray}
b=2\alpha \sqrt{t_c}, \quad t_c=\frac{T_c-T_{BKT}}{T_{BKT}},
\end{eqnarray}
where $\alpha$ determines the increase of the vortex-core energy, $\mu$, compared to the  XY-model valid for a single-layer superconductor: $\mu = \alpha \mu _{XY}$ \cite{Benfatto}. The fitting parameter $A$ was taken to be equal to 284. The fitting coefficient $b=0.274$, with $\rho_n=0.0698$ m$\Omega$cm, $T_{BKT}=92.60$~K and $T_c=92.74$~K (as taken from Fig.8(a)), gives $\alpha \simeq 3.5$. This value of $\alpha$ is similar to that derived in \cite{Benfatto2}.

Summarizing the experimental results we can state that both magnetization and resistivity measurements showed the 2D behavior of our YBCO single crystals close to $T_c$ and revealed a transition to the 3D state at lower temperatures. A characteristic non-monotonic temperature dependence of $R_c$ has been observed and interpreted as a result of the BKT transition. For one set of measurements, the BKT transition temperature $T_{BKT} = 92.60$~K was found from the $R_c(T)$ dependence (Fig. 8(a)) and the $\alpha(T)$ relation (Fig. 9(a)), and also derived from the BKT-AL interpolated formula which described correctly the $\rho_{ab}(T)$ behaviour at the superconducting transition (Fig. 8(a)). Thus, we believe that the existence of the BKT transition in our single crystals has been sufficiently well proved.

Experimental results on critical currents (Fig.9(b)) may be used for estimation of the possible error $\delta T$ in the determination of $T_c$ from the magnetic moment measurements in small magnetic fields. In the Meissner state, the screening current flows in a thin surface layer (Fig.10) with a thickness of about the penetration depth, which depends on the temperature as $\lambda_{ab} (T)= \lambda_{ab}(0)(1-T/T_c)^{-1/3}$, where $\lambda_{ab}(0) = 1.4\cdot 10^{-5}$~cm has been taken from Ref. \cite{Kamal}. Thus, the solid line in Fig.9(b) corresponds to the linear critical current density $j=I_c$(mA)$/0.5L$(cm)$=43[T_c(1-T/T_c)]^{2.33}/0.5L\simeq3.5\cdot10^4(1-T/T_c)^{2.33}$~A/cm, where $L=(0.090+0.098)/2=0.094$~cm is an average dimension of the single crystal along the main surface (see Fig.1(b)), and $T_c=92.74$~K. The $0.5L$ instead of $L$ was used assuming an approximate distribution of current in the single crystal. The magnetic moment of the sample with the main surface $S=3.2\cdot10^{-3}$~cm$^2$ is equal to $M(T)=0.1Sj(T)\lambda_{ab}(T)$~emu $\simeq 1.6\cdot10^{-4}(1-T/T_c)^2$~emu. For this estimation, the SQUID noise $M_S\simeq3\cdot 10^{-8}$~emu corresponds to $(1-T/T_c)\simeq1.4\cdot 10^{-2}$ or $\delta T\simeq1.3$~K for $T_c=92$~K. This estimation naturally explains the difference in the characteristic temperatures ($T_c$, $T_{BKT}$ and $T_{2D-3D}$) obtained from magnetic (Fig.4) and resistivity (Fig.7 and Fig.8) measurements.

Our results for the BKT transition have been obtained in conditions different from those reported earlier for YBCO single crystals \cite{Stamp,Yeh,Vas}. A single crystal investigated in \cite{Stamp} showed a wide resistive transition in the \textit{ab}-plane, $81.6-83.3$~K, and superconductivity along the \textit{c}-axis which appeared at around 76~K. $T_{BKT}=80$~K was obtained by fitting the BKT scaling to the temperature dependencies $R_{ab}(T)$ and $\alpha(T)$, where $\alpha$ is the exponent from the I-V characteristics. However, instead of $R_c(T)$ decreasing around $T_{BKT}$, its rise followed by a steep fall at lower temperatures was observed. This behaviour of the resistance along the \textit{c}-axis does not agree with the expected diminishing of $n_s$ (increasing of $n_n$) at $T\gtrsim T_{BKT}$. A single crystal investigated in \cite{Yeh} had a high critical temperature of about 93 K, but zero resistance in the \textit{ab}-plane and along the \textit{c}-axis occurred at the same temperature, i.e. superconductivity had a 3D character. For studies reported in \cite{Vas}, the resistance of a single crystal along the \textit{c}-axis was not measured, so the 2D character of the superconductivity was not proved. On the other hand, the BKT type transition has been confirmed in another cuprate, the stripe superconductor La$_{2-x}$Ba$_x$CuO$_4$ \cite{Li,Tranquada}.

The main problem that arises in measurements of I-V characteristics is the precise control of the sample temperature. Usually it is difficult to ensure very good thermal contact of a sample and a thermometer even when they are stuck together. Strongly non-linear current and the temperature dependence of the resistivity result in uncontrollable heating of the sample \cite{Mints} and the I-V characteristics may acquire even an S-shape form \cite{Lei}. Therefore, we believe that for unambiguous identification of the BKT transition in HTSC, simultaneous measurements of the resistivity both in the \textit{ab}-plane and along the \textit{c}-axis are necessary.

\section{Conclusions}

Magnetic measurements of YBCO single crystals revealed the 2D-3D transition in the superconducting state, in fields lower than 1.5~Oe. Magnetization curves obtained at temperatures above $T_{2D-3D}$ showed a two-stage transition to the normal state which allowed us to estimate the effective thickness $d_s\simeq1$ nm of a single quasi two-dimensional superconducting plate. The value of  $d_s$, which is important for the calculation of the helicity modulus (phase stiffness) of the superfluid phase \cite{Herbut,Boz}, is consistent with the crystal structure of YBCO containing the superconducting blocks of double CuO$_2$ planes with an yttrium layer.
Electric properties of YBCO single crystals were studied by simultaneous measurements of $\rho_{ab}$ and $R_c$ confirming the 2D-3D transition revealed by the magnetization results. Whereas the temperature dependence of $\rho_{ab}$ was typical in nature, $R_c(T)$ showed more complicated behavior, which was explained by the competition between a single-electron transport and Josephson tunneling. Our experiment showed that peculiarities of the $R_c(T)$ dependence might be used as a probe of characteristic changes in the superfluid density below the superconducting transition, where $\rho_{ab}=0$ is not informative. The Berezinskii-Kosterlitz-Touless state was observed in a narrow temperature range below the 2D mean-field superconducting transition.

\section*{Acknowledgments}

We are grateful to James Owen for important suggestions and critical comments regarding the final form of this article.



\bigskip
\begin{thebibliography}{W:99}
\bibitem{ref1}V. L. Berezinskii, Zh. Eksp. Teor. Fiz.  \textbf{59}, 907 (1970) [Sov. Phys. JETP \textbf{32}, 493 (1971)].
\bibitem{ref2}V. L. Berezinskii, Zh. Eksp. Teor. Fiz. \textbf{61}, 1144 (1971) [Sov. Phys. JETP \textbf{34}, 610–616 (1972)].
\bibitem{ref3}J. M. Kosterlitz and D. J. Thouless, Journal of Physics C: Solid State Physics \textbf{6}, 1181–1203 (1973).
\bibitem{Min}P. Minnhagen, Rev. Mod. Phys. \textbf{59}, 1001 (1987).
\bibitem{Halperin}B. I. Halperin and D. R. Nelson, Journal of Low Temperature Physics \textbf{36}, 599 (1979).
\bibitem{Benfatto}L. Benfatto, C. Castellani, and T. Giamarchim, Phys. Rev. B \textbf{80}, 214506 (2009).
\bibitem{exper}C. Xu, L. Wang, Z. Liu, L. Chen, J. Guo, N. Kang, X.-L. Ma,
H.-M. Cheng, and W. Ren, Nature Mat. \textbf{14}, 1135 (2015).
\bibitem{Rep}J. M. Repaci, C. Kwon, Qi Li, Xiuguang Jiang, T. Venkatessan, R. E. Glover III, C. J. Lobb, and R. S. Newrock,  Phys. Rev. B \textbf{54}, R9674 (1996).
\bibitem{Kogan}V. G. Kogan, Phys. Rev. B \textbf{75}, 064514 (2007).
\bibitem{AL}L. G. Aslamasov and A. I. Larkin, Phys. Lett. A \textbf{26}, 238 (1968).
\bibitem{Wang1}L. Li, Y. Wang, M.J. Naughton, S. Ono, Y. Ando, and N. P. Ong, Europhys. Lett., \textbf{72}, 451–457 (2005).
\bibitem{Wang2}Y. Wang, L. Li, M. J. Naughton, G. Gu, and N. P. Ong, Phys. Rev. Lett. \textbf{95}, 247002 (2005).
\bibitem{Oganesyan}V. Oganesyan, D. A. Huse, and S. L. Sondhi, Phys. Rev. B \textbf{73}, 094503 (2006).
\bibitem{Wang3}L. Li, Y. Wang, S. Komiya, S. Ono, Y. Ando, G. D. Gu, and N. P. Ong, Phys. Rev. B \textbf{81}, 054510 (2010).
\bibitem{ref11} D. N. Astrov, N. B. Ermakov, A. S. Borovik-Romanov, E. G. Kolevatov, and V. I. Nizhankovskii, Pis’ma Zh. E´ksp. Teor. Fiz. \textbf{63}, 713 (1996) [JETP Letters \textbf{63}, 745 (1996)].
\bibitem{ref12}V. I. Nizhankovskii, A. I. Khar'kovskii, and A. J. Zaleski,
Eur. Phys. J. B \textbf{10}, 761 (1999).
\bibitem{ref13}V. I. Nizhankovskii, JMMM \textbf{242-245}, 928 (2002).
\bibitem{Pauw}Van der Pauw, Philips Res. Repts \textbf{13}, 1 (1958).
\bibitem{Sch1}P. Schnabel, Philips Res. Repts \textbf{19}, 43 (1964).
\bibitem{Sch2}P. Schnabel, Z. angew. Phys. \textbf{22}, 136 (1967).
\bibitem{Lug}L. B. Luganskii and V. I. Tsebro, Instrum. Exp. Tech. \textbf{58}, 118 (2015); in Eq.(16) the number $\pi$ has been omitted (see: arXiv:1502.02600 [cond-mat.mtrl-sci]).
\bibitem{Nag}K. Nagasao, T. Masui and S. Tajima, Physica C \textbf{468}, 1188 (2008).
\bibitem{Shams}G. A. Shams, J. W. Cochrane, and G. J. Russell, Physica C \textbf{363}, 243 (2001).
\bibitem{Uch}H. Uchiyama, N. Matsukura, and N. Chikumoto, Phys. Rev. B  \textbf{81}, 060511(R) (2010).
\bibitem{Vovk}R. V. Vovk, N. R. Vovk, O. V. Shekhovtsov, I. L. Goulatis, and
A. Chroneos, Supercond. Sci. Technol. \textbf{26}, 085017 (2013).
\bibitem{Gray}K. E. Gray and D. H. Kim, Phys. Rev. Lett. \textbf{70}, 1693 (1993).
\bibitem{ref14}M. Tinkham, \textit{Introduction to Superconductivity} (Dover, New York, 2004).
\bibitem{ref15}J. Kurkij\"{a}rvi and V. Ambegaokar, Phys. Rev. B \textbf{5}, 868 (1972).
\bibitem{Lambda_c}T. Schneider, J. Hofer, M. Willemin, J.M. Singer, and H. Keller, Eur. Phys. J. B \textbf{3}, 413 (1998).
\bibitem{Schneider}T. Schneider and J. M. Singer, \textit{Phase Transition Approach to
High Temperature Superconductivity} (Imperial College Press, London, 2000), Chap. 6.
\bibitem{Bul}L. N. Bulaevskii, M. Ledvij, and V. G. Kogan, Phys. Rev. Lett. \textbf{68}, 3773 (1992).
\bibitem{Kosh}A. E. Koshelev, Phys. Rev. B \textbf{50}, 506 (1994).
\bibitem{Jun}A. Junod, J.-Y. Genoud, G. Triscone, and T. Schneider, Physica C \textbf{294}, 115 (1998).
\bibitem{Benfatto2}L. Benfatto, C. Castellani, and T. Giamarchi, Phys. Rev. Lett. \textbf{98}, 117008 (2007).
\bibitem{Kamal}S. Kamal, Ruixing Liang, A. Hosseini, D. A. Bonn, and W. N. Hardy, Phys. Rev. B \textbf{58}, R8933 (1998).
\bibitem{Stamp}P. C. E. Stamp, L. Forro, and C. Ayache, Phys. Rev. \textbf{38}, 2847 (1988).
\bibitem{Yeh}N.-C. Yeh and C. C. Tsuei, Phys. Rev. B \textbf{39}, 9708 (1989).
\bibitem{Vas}M. A. Vasyutin, A. I. Golovashkin, and N. D. Kuz’michev, Physics of Solid State \textbf{48}, 2260 (2006).
\bibitem{Li}Q. Li, M. H\"{u}cker, G. D. Gu, A. M. Tsvelik, and J. M. Tranquada, Phys. Rev. Lett. \textbf{99}, 067001 (2007).
\bibitem{Tranquada}J.M. Tranquada, G. D. Gu, M. H\"{u}cker, Q. Jie, H.-J. Kang, R. Klingeler, Q. Li, N. Tristan, J. S. Wen, G. Y. Xu, Z. J. Xu, J. Zhou, and M. v. Zimmermann, Phys. Rev. B \textbf{78}, 174529 (2008).
\bibitem{Mints}A. Vl. Gurevich and R. G. Mints, Rev. Mod. Phys. \textbf{59}, 941 (1987).
\bibitem{Lei}L. Qiao, D. Li, S. V. Postolova, A. Yu. Mironov, V. Vinokur, and B. Rosenstein, Scientific Reports \textbf{8}, 14104 (2018).
\bibitem{Herbut}I. F. Herbut and M. J. Case, Phys. Rev. B \textbf{70}, 094516 (2004).
\bibitem{Boz}I. Bo$\check{z}$ovi$\acute{c}$, X. He1, J. Wu1, and A. T. Bollinger, Nature \textbf{536}, 309 (2016).
\end{thebibliography}
\end{document}